\documentclass[preprint,12pt]{elsarticle}

\usepackage{amssymb}
\usepackage[utf8]{inputenc}
\usepackage[T1]{fontenc}
\journal{NIM A}

\begin{document}

\begin{frontmatter}

%% Title, authors and addresses

%% use the tnoteref command within \title for footnotes;
%% use the tnotetext command for theassociated footnote;
%% use the fnref command within \author or \address for footnotes;
%% use the fntext command for theassociated footnote;
%% use the corref command within \author for corresponding author footnotes;
%% use the cortext command for theassociated footnote;
%% use the ead command for the email address,
%% and the form \ead[url] for the home page:
%% \title{Title\tnoteref{label1}}
%% \tnotetext[label1]{}
%% \author{Name\corref{cor1}\fnref{label2}}
%% \ead{email address}
%% \ead[url]{home page}
%% \fntext[label2]{}
%% \cortext[cor1]{}
%% \affiliation{organization={},
%%             addressline={},
%%             city={},
%%             postcode={},
%%             state={},
%%             country={}}
%% \fntext[label3]{}

\title{CdZnTe detectors tested at the $DA\Phi NE$ collider for future kaonic atoms measurements}

%% use optional labels to link authors explicitly to addresses:
%% \author[label1,label2]{}
%% \affiliation[label1]{organization={},
%%             addressline={},
%%             city={},
%%             postcode={},
%%             state={},
%%             country={}}
%%
%% \affiliation[label2]{organization={},
%%             addressline={},
%%             city={},
%%             postcode={},
%%             state={},
%%             country={}}

\author[1]{A. Scordo\corref{cor1}}\ead{alessandro.scordo@lnf.infn.it}
\author[2]{L. Abbene}
\author[1,4]{F. Artibani}
\author[1]{M. Bazzi}
\author[5]{M. Bettelli}
\author[6]{D. Bosnar}
\author[7]{G. Borghi}
\author[8]{M. Bragadireanu}
\author[2]{A. Buttacavoli}
\author[3]{M. Cargnelli}
\author[7]{M. Carminati}
\author[1]{A. Clozza}
\author[1,4]{F. Clozza}
\author[1]{L. De Paolis}
\author[7]{G. Deda}
\author[9,1]{R. Del Grande}
\author[9]{L. Fabbietti}
\author[7]{C. Fiorini}
\author[8]{I. Fri\v{s}\v{c}i\'{c}}
\author[1]{C. Guaraldo}
\author[1]{M. Iliescu}
\author[10]{M. Iwasaki}
\author[1]{A. Khreptak}
\author[1]{S. Manti}
\author[3]{J. Marton}
\author[11,12]{P. Moskal}
\author[1]{F. Napolitano}
\author[11,12]{S. Nied\'zwiecki}
\author[13]{H. Ohnishi}
\author[14,1]{K. Piscicchia}
\author[2]{F. Principato}
\author[13]{Y. Sada}
\author[1]{F. Sgaramella}
\author[3]{H. Shi}
\author[11,12]{M. Silarski}
\author[1,14,7]{D. L. Sirghi}
\author[1,8]{F. Sirghi}
\author[11,12]{M. Skurzok}
\author[1]{A. Spallone}
\author[13]{K. Toho}
\author[3]{M. T\"uchler}
\author[13]{C. Yoshida}
\author[5]{A. Zappettini}
\author[3]{J. Zmeskal}
\author[1]{C. Curceanu}

\cortext[cor1]{Corresponding author}

\affiliation[1]{organization={Laboratori Nazionali di Frascati, INFN}, addressline={Via E. Fermi 54}, city={Frascati}, postcode={00044}, country={Italy}}
\affiliation[2]{organization={Dipartimento di Fisica e Chimica - Emilio Segrè, Università di Palermo}, addressline={Viale Delle Scienze, Edificio 18}, city={Palermo}, postcode={90128}, country={Italy}}
\affiliation[3]{organization={Stefan-Meyer-Institut f\"ur subatomare Physik}, addressline={Dominikanerbastei 16}, city={Vienna}, postcode={1010}, country={Austria}}
\affiliation[4]{organization={Dipartimento di Fisica dell’Universita` “Sapienza”}, addressline={Piazzale A. Moro 2}, city={Rome}, postcode={00185}, country={Italy}}
\affiliation[5]{organization={Istituto Materiali per l’Elettronica e il Magnetismo, Consiglio Nazionale delle Ricerche}, addressline={Parco Area delle Scienze 37/A}, city={Parma}, postcode={43124}, country={Italy}}
\affiliation[6]{organization={Department of Physics, Faculty of Science , University of Zagreb}, addressline={Bijeni\v cka cesta 32}, city={Zagreb}, postcode={10000}, country={Croatia}}
\affiliation[7]{organization={Dipartimento di Elettronica, Informazione e Bioingegneria and INFN Sezione di Milano, Politecnico di Milano}, addressline={Via Giuseppe Ponzio 34}, city={Milano}, postcode={20133}, country={Italy}}
\affiliation[8]{organization={Horia Hulubei National Institute of Physics and Nuclear Engineering (IFIN-HH)}, addressline={No. 30, Reactorului Street}, city={Magurele, Ilfov}, postcode={077125}, country={Romania}}
\affiliation[9]{organization={Physik Department E62, Technische Universit\"at M\"unchen}, addressline={James-Franck-Stra{\ss}e 1}, city={Garching}, postcode={85748}, country={Germany}}
\affiliation[10]{organization={Institute of Physical and Chemical Research, RIKEN}, addressline={2-1 Hirosawa}, city={Wako, Saitama}, postcode={351-0198}, country={Japan}}
\affiliation[11]{organization={Faculty of Physics, Astronomy, and Applied Computer Science, Jagiellonian University}, addressline={ul. prof. Stanis\l awa \L ojasiewicza 11}, city={Krak\'ow}, postcode={30-348}, country={Poland}}
\affiliation[12]{organization={Center for Theranostics, Jagiellonian University}, addressline={ul. prof. Stanis\l awa \L ojasiewicza 11}, city={Krak\'ow}, postcode={30-348}, country={Poland}}
\affiliation[13]{organization={Research Center for Electron Photon Science (ELPH), Tohoku University}, addressline={1-2-1 Mikamine}, city={Taihaku-ku, Sendai, Miyagi}, postcode={982-0826}, country={Japan}}
\affiliation[14]{organization={Centro Ricerche Enrico Fermi-Museo Storico della fisica e Centro Studi e Ricerche “Enrico Fermi”}, addressline={Via Panisperna, 89a}, city={Rome}, postcode={00184}, country={Italy}}

\begin{abstract}
%% Text of abstract

The SIDDHARTA-2 collaboration at the INFN Laboratories of Frascati (LNF)
aims to perform groundbreaking measurements on kaonic atoms. In parallel
and beyond the ongoing kaonic deuterium, presently running on the DA$\Phi$NE
collider at LNF, we plan to install additional detectors to perform further
kaonic atoms’ studies, taking advantage of the unique low energy and low
momentum spread $K^-$ beam delivered by the at-rest decay of the $\phi$ meson.

\noindent CdZnTe devices are ideal for detecting transitions toward both the upper and
lower levels of intermediate-mass kaonic atoms, like kaonic carbon and aluminium, which have an important impact on the strangeness sector of nuclear
physics.

\noindent We present the results obtained in a set of preliminary tests conducted on
DA$\Phi$NE, in view of measurements foreseen in 2024, with the twofold aim to
tune the timing window required to reject the extremely high electromagnetic
background, and to quantify the readout electronics saturation effect due to
the high rate, when placed close to the Interaction Region (IR). In the first
test we used commercial devices and electronics, while for the second one
both were customized at the IMEM-CNR of Parma and the University of
Palermo.

\noindent The results confirmed the possibility of finding and matching a proper timing
window where to identify the signal events and proved better performances,
in terms of energy resolution, of the custom system. In both cases, strong
saturation effects were confirmed, accounting for a loss of almost 90\% of the
events, which will be overcome by a dedicated shielding structure foreseen
for the final experimental setup.

\end{abstract}

%%Graphical abstract
%\begin{graphicalabstract}
%\includegraphics{grabs}
%\end{graphicalabstract}

%%Research highlights
%\begin{highlights}
%\item Research highlight 1
%\item Research highlight 2
%\end{highlights}

\begin{keyword}
%% keywords here, in the form: keyword \sep keyword
Kaonic Atoms \sep CdZnTe detectors \sep GEANT4 simulations \sep strangeness nuclear physics \sep Kaon-nucleon(s) interaction
%% PACS codes here, in the form: \PACS code \sep code

%% MSC codes here, in the form: \MSC code \sep code
%% or \MSC[2008] code \sep code (2000 is the default)

\end{keyword}

\end{frontmatter}

%% \linenumbers

%% main text
\section{Introduction}
\label{sec:intro}

\noindent The energies of the radiative transitions of kaonic atoms, and in particular of those towards their innermost levels where the strong interaction is not negligible,
represent ingredients on which the present knowledge on low energy strangeness nuclear physics is built \cite{RevModPhys.91.025006}.
They have been widely investigated in the 1970's and 1980's, when a systematic study of such systems for a wide set of atoms along the periodic table has been performed. This allowed to setup a dataset which served, until nowadays, as the main experimental input for the development of many theoretical models \cite{Davies:1979,Izycki:1980,Bird:1983,Wiegand:1971zz,Baird:1983ub,Friedman:1994hx}.

\noindent The energies and the widths of such transitions span over a very wide range, going from a few keV up to almost 1 MeV, depending on the Z of the selected atom and on the two levels among which the transition occurs.
The choice of the proper detector to measure the kaonic atoms of interest is then crucial, and it has to be taken as a function of many factors, like their efficiency in a given energy range, an energy resolution able to resolve the lines of interest with their natural linewidths, excellent linearity, well-known calibration function,
and the capability to work in typically high background environments like particle accelerators, where kaons are produced.

\noindent The most recent results on light kaonic atoms, however, shed a shadow on the aforementioned database, since they are not reproducing the results obtained in the past \cite{RevModPhys.91.025006,Iwasaki:1997,Okada:2007ky,SIDDHARTA:2011dsy,SIDDHARTA:2009qht,SIDDHARTA:2010uae,SIDDHARTA:2012rsv}. Furthermore, many of the earliest measurements of the strong interaction-induced 
shift and width of the lower levels in atoms like carbon, aluminium and sulfur suffer from very large uncertainties and, in some cases, they are hardly reproducible with the most recent theoretical models \cite{Obertova:2022wpw}.
All these facts renewed the interest in performing a systematic study of kaonic atoms, and the DA$\Phi$NE collider at LNF \cite{Zobov:2010zza,Milardi:2021khj}, with its unique low momentum and low dispersion kaon beam, is the most suitable facility in the world to pursue this scope.

\noindent The SIDDHARTA-2 collaboration aims to perform, in parallel with the measurement of kaonic deuterium at DA$\Phi$NE, additional measurements on intermediate-mass kaonic atoms like KC, KAl and KS, strongly demanded by the theoretical community.
The transitions of interest for such systems lay in the 30-300 keV range, where CdZnTe devices are the radiation detectors best fulfilling the high efficiency and high-resolution requirements demanded to measure 
the emitted X-rays with precisions of a few tens of eV, thus pinning down those of the older experiments. Moreover, with their excellent performances at room temperatures, CdZnTe detectors allow for realizing small and compact detection systems,
easy and fast to be installed and integrated with the existing SIDDHARTA-2 apparatus.
The behaviour of CdZnTe devices in high electromagnetic background environments is not, however, well known and ad-hoc crystals and detectors have to be realized to mantain a high efficiency in the
whole energy range, without worsening the spectroscopic resolution. Also, it is crucial to define a perfect timing window for the data acquisition to reject a huge amount of background events.

\noindent The SIDDHARTA-2 experiment already performed, in 2022, preliminary tests in this direction \cite{Abbene:2023ogi,Abbene:2023vhm} with a small ($1\,cm^2$) CdZnTe detector installed 45 cm away from the IR. Very promising results in terms of linearity, resolution and high-rate capability were obtained, which triggered the realization of a bigger
experimental setup for the physics measurement to be performed in 2024.

\noindent In this paper, we present the outcomes of two additional tests performed in the first half of 2023 with CdZnTe devices installed at a few tens of cm from the Interaction Point, aiming at tuning the time selection window for the signal identification and investigating the impact of the high electromagnetic background.
\section{Experimental setup}
\label{sec:setup}

\noindent Kaonic atoms are formed when a $K^-$, produced in a back-to-back configuration with a $K^+$ from the $\Phi$ decays, is stopped in a selected target; subsequently, the X-rays isotropically emitted during the atomic cascade process \cite{Koike:1998uz,Kalantari:2012isg,Berezin:1970qw} can be measured by the radiation detector.
However, the background of the DA$\Phi$NE machine plays a crucial role when the detector needs to be placed close to the Interaction Region (IR), where
rates up to $\mathrm{MHz/cm^2}$ level can be reached.
The main components of this background are two: a first one, synchronous with the kaonic atoms signal coming from Minimum Ionized Particles (MIPs) produced within the $e^+e^-$ collision events and releasing energy in the detector, and a second one, asynchronous with the signal and mostly due to particles lost from the beam by Touscheck and beam-gas effects.
The hardware rejection of these backgrounds is crucial for the success of the measurements.

\noindent To this scope, we plan to exploit the already existing SIDDHARTA-2 Luminosity Monitor system (LM), consisting of two plastic scintillators placed a few cm away from the
IR, installed in the SIDDHARTA-2 apparatus at DA$\Phi$NE to assess the integrated luminosity \cite{Skurzok:2020phi}.
The LM works in coincidence with the DA$\Phi$NE radiofrequency clock, so that a trigger from such a system is a clear signature of an event occurring in time with a collision.

\begin{figure}[h]%
\centering
	\includegraphics[width=1.\textwidth]{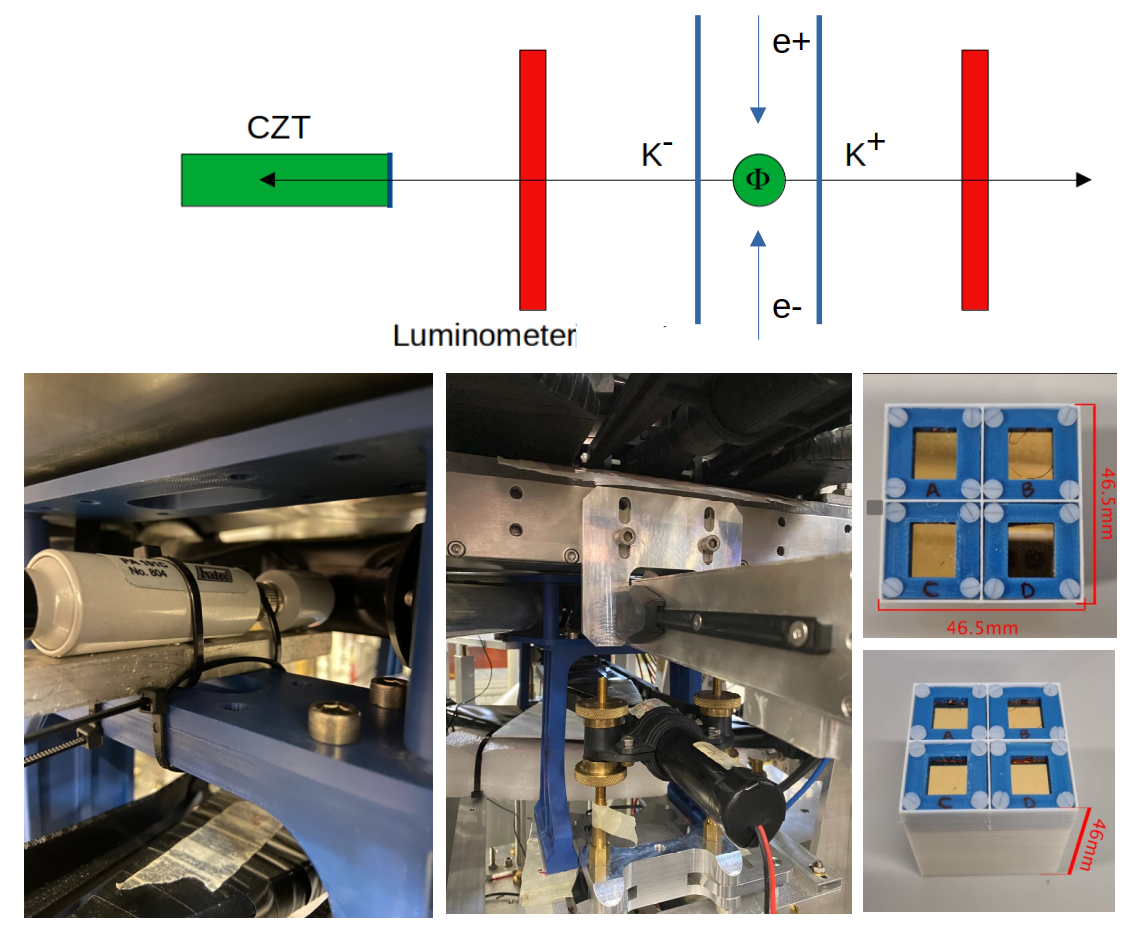}
	\caption{\emph{Top: Schematic of the main components for the two CdZnTe tests (not in scale)}  \emph{Bottom: picture of the installed RITEC CZT/500 detector (left), of the custom one (center) and of the unit cell installed in its aluminum box (right)}.}\label{fig:setup_scheme-pictures}
\end{figure}

\noindent An additional Time-Of-Flight analysis allows to clearly distinguish between MIPs, mostly coming from the electromagnetic showers created in the magnets before the IR, and a $K^+K^-$ pair. 
Furthermore, the readout of the CdZnTe detectors needs to be performed within a very specific timing window, triggered by the SIDDHARTA-2 LM system, to be carefully tuned.
To do so, we performed tests placing CdZnTe detectors after the LM scintillator in the anti boost direction of the DA$\Phi$NE machine \cite{Skurzok:2020phi}.
A schematic of the setup is shown in the upper part of Fig. \ref{fig:setup_scheme-pictures}.

\noindent The first test was conducted with a RITEC CZT/500 detector system, a quasi-hemispherical CdZnTe device of 0.5 $\mathrm{cm^3}$ active volume equipped with a commercial 
PA101C RITEC preamplifier; the analogue signals have been acquired with a CAEN DT5780 Digitizer in coincidence with the SIDDHARTA-2 LM ones, preliminarily
processed by a Time-to-Analogue Converter (TAC) from ORTEC, similar to our previous work \cite{Abbene:2023ogi,Abbene:2023vhm}.
The CZT/500 was installed immediately after the SIDDHARTA-2 LM, 7.5 cm away from the Interaction Region.

\noindent The second test was conducted using four custom $\mathrm{13\times15\times5\,mm^3}$ quasi-hemispherical CdZnTe detectors provided by Redlen Technologies (Saanichton, BC, Canada), securely housed in PLA holders at IMEM-CNR (Parma) and wire-bonded to the analogue charge-sensitive preamplifiers (CSPs) and digital pulse processing (DPP) based readout electronics. 
The four channels of the detection system were AC coupled to hybrid CSPs, developed at DiFC of the University of Palermo, and processed by a 64-channel DPP electronics. The preamplifiers were characterized by an equivalent noise charge (ENC) of about 100 electrons (equivalent to about 1 keV FWHM for CZT detectors) and equipped with a resistive feedback circuit with exponential decay and time constant of 100 µs. The digital electronics consisted of a CAEN V2740 digitizer driven by an original firmware developed at the University of Palermo as well.

\noindent Both the detectors and the preamplifiers were enclosed in a 5 mm thick aluminium box with an entrance window of $\mathrm{0.2\,mm}$.
The whole apparatus was installed in a position corresponding to a distance between the detectors and the IR of 22 cm.

\noindent In the lower part of Fig. \ref{fig:setup_scheme-pictures}, a picture of the installed commercial RITEC CZT/500 detector (left), of the custom one (centre) and of the unit cell installed in its aluminum box (right) can be seen.

\section{Results}
\label{sec:results}

\begin{figure}[h]%
\centering
	\includegraphics[width=1.\textwidth]{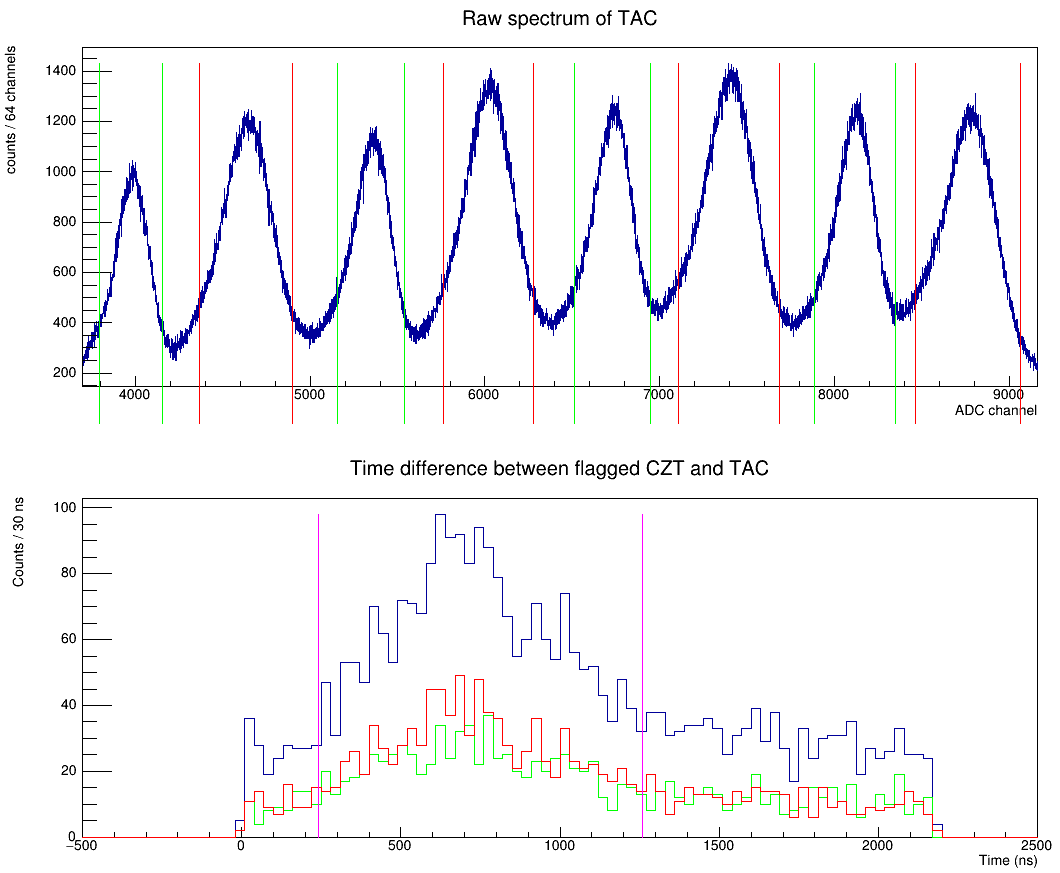}
	\caption{\emph{Top: SIDDHARTA-2 LM TOF spectrum, where windows for Kaons (green) and MIPs (red) time selection are highlighted.}  
	\emph{Bottom: Timing spectrum of the CdZnTe with respect to the SIDDHARTA-2 LM (blue) with further selection of Kaons (green) and MIPs (red).}}\label{fig:1cm2test_plots}
\end{figure}

\noindent The goal of the first test was to tune the correct timing window for the CdZnTe signals' acquisition in coincidence with the SIDDHARTA-2 LM. The main results
are presented in Fig. \ref{fig:1cm2test_plots}.
In the upper pad, we show the TOF spectrum obtained from the processing of the LM signals through the TAC module. The 8 peaks structure, as well as the acquisition logic,
have been described in \cite{Abbene:2023ogi,Abbene:2023vhm}; here, the four peaks within the green lines correspond to $K^-$ interacting in the plastic scintillators, while those within the red ones 
are produced by MIPs.
In the lower pad of Fig. \ref{fig:1cm2test_plots}, the time distribution of the CdZnTe events occurring in coincidence with a LM signal is shown in blue, while in red and green 
the same quantity is shown with the additional kaons (green) or MIPs(red) selection.
These distributions show how, within a 1 $\mu s$ timing window, a clear peak of events occurring in time with the LM can be found. It is worth reminding here that 
in this specific test, where no target where to stop the kaons was present, both kaons and MIPs reached the detector and could give a signal as clearly visible by the two coloured distributions.

\noindent The evidence of a time-coincidence peak is a very important result since it confirms the good timing capabilities of CdZnTe and the possibility of using them in kaonic atom measurements.
Another key parameter to monitor was the saturation of the readout electronics induced by the very high rate on the detector. 
In the first test, due to the limitations of the electronics, the saturating events were removed hardware, 
preventing a proper quantification.
During the second test, performed with the custom detection system, such quantification was instead allowed by the new front-end and readout electronics realized in Palermo and by
the larger distance from the IR. A quantitative analysis performed on the acquired data revealed how almost 90\% of the events recorded by the CdZnTe detectors were saturating
the electronics; this was a very well-known issue which will be overcome in the final setup employing of a proper shielding system.

\begin{figure}[h]%
\centering
	\includegraphics[width=1.\textwidth]{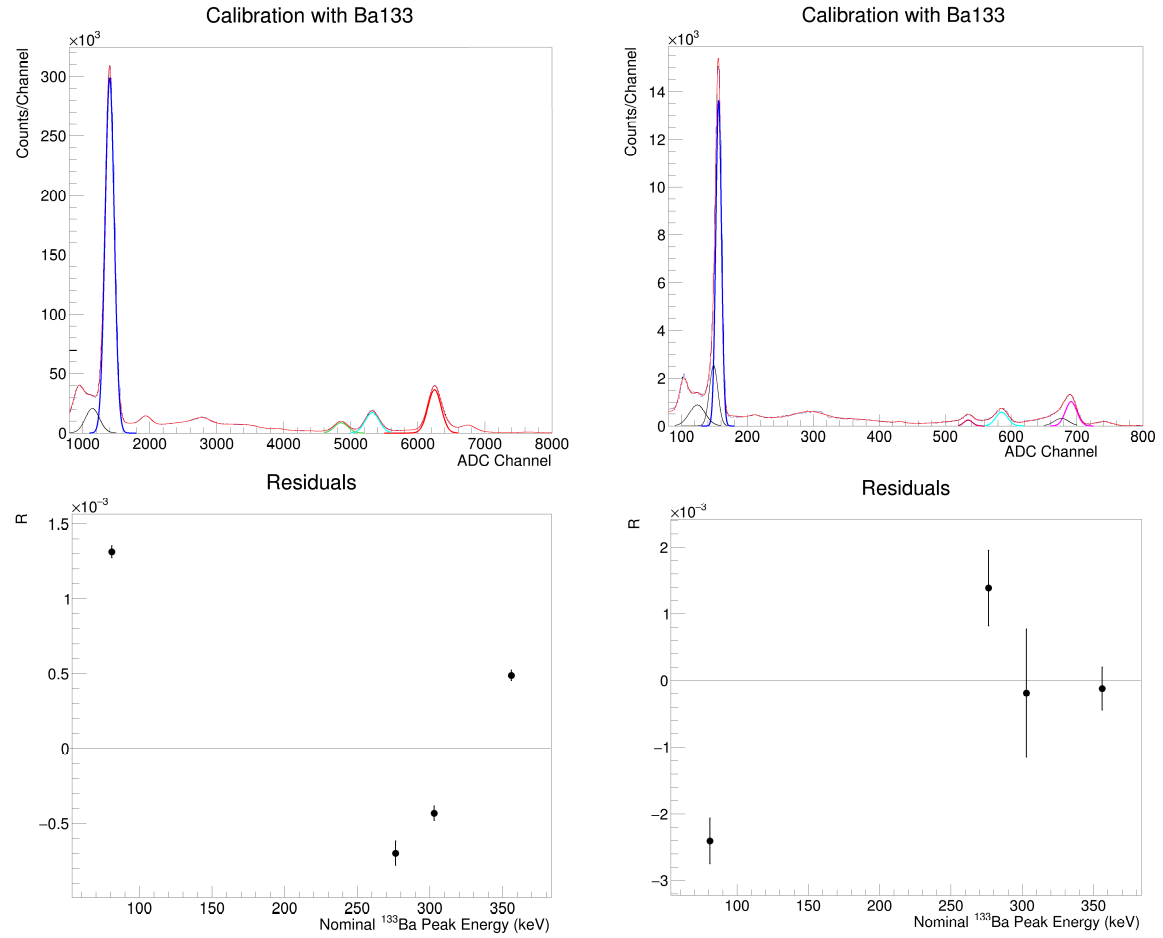}
	\caption{\emph{Top: fitted energy spectra are shown for the CZT/500 (left) and for one of the custom detectors (right). 
	The gaussians corresponding the four main peaks used for the energy calibration are shown in colours, while the auxiliary ones are shown in black (see text).}  
	\emph{Bottom: relative deviations of the mean energy values obtained from a fit on the energy calibrated spectra and the nominal ones for the CZT/500 (left) and the 
	custom CdZnTe system (right).}}\label{fig:compare_res}
\end{figure}

\noindent During both tests, the detectors were calibrated with a $^{133}Ba$ source, and their linearity and energy resolution were measured.
In the upper part of Fig. \ref{fig:compare_res}, the fitted energy spectra are shown for the CZT/500 (left) and for one of the custom detectors (right).
The linear calibration was obtained using four of the $^{133}Ba$ characteristic lines, for which the individual gaussians resulting from the fit are shown in colour together
with the overall fitting function. Some of the peaks showed a more evident Compton tail, which was taken into account including additional gaussians, at lower energies with respect
to the main one, reported in black. In the case of the custom detector, an additional characteristic peak at 79.6 keV is visible, also shown in black, while in the commercial one the worse energy resolution prevented it from distinguishing it.
The residuals (R) shown in the lower pads of Fig. \ref{fig:compare_res} represent the relative deviations of the positions of the four peaks obtained from a fit of the energy-calibrated spectra with respect to the nominal ones for the commercial (left) and custom (right) system; such residuals reveal a linearity of the systems better than $\mathrm{2\times10^{-3}}$. 

\noindent As an additional result, we report how the energy resolutions obtained with the custom detection system are better than those obtained with the commercial one,
confirming the good performances of both the detectors and the frontend electronics realized in Parma and Palermo.
A summary of the obtained results can be found in Tab. \ref{tab:summary}.

\begin{table}[h]
\centering
  \begin{tabular}{ | c | c | c | c | c |}
    \hline
    & \multicolumn{2}{|c|}{RITEC CZT/500} & \multicolumn{2}{|c|}{IMEM-CNR CdZnTe} \\ \hline 
	  $\mathrm{^{133}Ba}$ Peak (keV) & R ($10^{-3}$) & FWHM/E  & R ($10^{-3}$) & FWHM/E \\ \hline
	  81.0 & 	 1.3	& 0.110	& -2.4 & 0.064\\ \hline 
	  276.4 &	 -0.7	& 0.048	& 1.4	& 0.028\\ \hline 
	  302.9 &	 -0.4	& 0.046	& -0.2	& 0.036\\ \hline 
	  356.0 &	 0.5	& 0.037	& -0.1	& 0.030\\ \hline 
  \end{tabular}
	\caption{Summary of the results obtained with the two systems employed during the tests (see text for more details).}
  \label{tab:summary}
\end{table}

\section{Conclusions}
\label{sec:conc}

\noindent In this paper, we reported the results of two preliminary tests conducted with two different CdZnTe detection systems on the DA$\Phi$NE collider at INFN Laboratories of Frascati,
aiming at assessing the saturation effect due to the high machine background and at tuning the timing window for the acquisition of the detector signal in coincidence with the trigger given by the SIDDHARTA-2 Luminosity Monitor.
These tests revealed how, within a 1 $\mu s$ timing window, a clear peak of events occurring in time with the LM can be found; this represents a crucial test proving how 
the good timing resolution of such devices allows to employ them in kaonic atoms measurements.
Finally, we demonstrated that the energy resolutions obtained with the custom detection system are better than those obtained with the commercial one,
confirming the good performances of both the detectors and the frontend electronics realized in Parma and Palermo.
Dedicated measurements of selected kaonic atoms like KC, KAl and KS, will be performed in 2024 by the SIDDHARTA-2 collaboration on the DA$\Phi$NE collider, in parallel with the ongoing measurement of the kaonic deuterium.

\section*{Acknowledgements}
\noindent 
We thank C. Capoccia from LNF-INFN and H. Schneider, L.
Stohwasser, and D. Pristauz-Telsnigg from Stefan Meyer-Institut for
their fundamental contribution in designing and building the
SIDDHARTA-2 setup. We thank as well the DA$\Phi$NE staff for
the excellent working conditions and permanent support. 
We thank C. Capoccia from LNF-INFN and H. Schneider, L. Stohwasser, and D. Pristauz- Telsnigg from Stefan-
Meyer-Institut for their fundamental contribution in designing and building the SIDDHARTA-2 setup. We thank as
well the $\mathrm{DA\Phi NE}$ staff for the excellent working conditions and permanent support. Part of this work was supported by 
the EU STRONG-2020 (JRA8) project within HORIZON 2020, Grant agreement ID: 824093; the Austrian Science Fund (FWF): [P24756-N20 and P33037-N]; 
the Croatian Science Foundation under the project IP-2018-01-8570; Japan Society for the Promotion of Science JSPS KAKENHI Grant No. JP18H05402; 
the Polish Ministry of Science and Higher Education Grant No. 7150/E-338/M/2018 and the Polish
National Agency for Academic Exchange (Grant no PPN/BIT/2021/1/00037); the SciMat and qLife Priority Research Areas budget under 
the program Excellence Initiative - Research University at the Jagiellonian University.
Catalina Curceanu acknowledge University of Adelaide, where part of this work was done (under the George Southgate
fellowship, 2023).

%% The Appendices part is started with the command \appendix;
%% appendix sections are then done as normal sections
%% \appendix

%% \section{}
%% \label{}

%% If you have bibdatabase file and want bibtex to generate the
%% bibitems, please use
%%
  \bibliographystyle{elsarticle-num} 
  \bibliography{channeling2023}

%% else use the following coding to input the bibitems directly in the
%% TeX file.

%\begin{thebibliography}{00}

%% \bibitem{label}
%% Text of bibliographic item

%\bibitem{}

%\end{thebibliography}
\end{document}